\documentclass[superscriptaddress,showpacs,preprintnumbers,amsmath,amssymb]{revtex4}

\usepackage{graphicx}
\usepackage{dcolumn}
\usepackage{epsfig}
\usepackage{bm}

\def\eq#1{{(\ref{#1})}}
\def\fig#1{{Fig.~\ref{#1}}}

\newcommand{\beq}{\begin{equation}}
\newcommand{\eeq}{\end{equation}}
\newcommand{\ben}{\begin{eqnarray*}}
\newcommand{\een}{\end{eqnarray*}}

\newcommand{\as}{\alpha_s} 
\newcommand{\un}{\underline}   

\begin{document}

\preprint{BNL--NT--04/16}
\preprint{NT@UW--04--010}

\title{Nuclear Modification Factor in d+Au Collisions: \\
Onset of Suppression in the Color Glass Condensate}

\author{Dmitri Kharzeev}
\affiliation{
Physics Department, P. O. Box 500 \\ Brookhaven National Laboratory \\
Upton, NY 11973 }

\author{Yuri V. Kovchegov}
\affiliation{Department of Physics, Box 351560 \\
University of Washington \\
Seattle, WA 98195-1560
}

\author{Kirill Tuchin}
\affiliation{
Physics Department, P. O. Box 500 \\ Brookhaven National Laboratory \\
Upton, NY 11973 }

\date{\today}

\begin{abstract}
We perform a quantitative analysis of the nuclear modification factor in
deuteron--gold collisions $R^{dAu}$ within the Color Glass Condensate 
approach, and compare our results with the recent data from RHIC 
experiments.  Our model leads to Cronin enhancement at
mid-rapidity, while at forward rapidities it predicts strong
suppression of $R^{dAu}$ at all $p_T$ due to low-$x$ evolution. We
demonstrate that our results are consistent with the data for $dAu$
charged hadron spectra, $R^{dAu}$ and $R^{CP}$ recently reported for
rapidities in the interval $\eta =0 \div 3.2$ by the BRAHMS experiment at
RHIC. We also make a prediction for $R^{pA}$ at mid-rapidity in $pA$ collisions 
at the LHC.

\end{abstract}



\maketitle
Recent observations \cite{Debbe,BRAHMSdata,RHICres2,RHICres3,RHICres4}
of the suppression of high $p_T$ hadron yields at forward rapidities
at RHIC have attracted considerable interest. The observed suppression
is in sharp contradiction with the naive multiple scattering picture,
in which the magnitude of Cronin enhancement observed at mid-rapidity
is expected to increase further at forward rapidities, reflecting the
growth of the number of scattering centers (partons) at small Bjorken
$x$. On the other hand, the observed effect has been predicted
\cite{KLM,KKT,Alb,BKW} as a signature of quantum evolution in the
Color Glass Condensate (CGC) \cite{GLR,mq,BM,MV,kjklw,BK,JKLW}. Very
recently, the first exploratory experimental results \cite{STAR-Les}
on the back--to--back azimuthal correlations of high $p_T$ particles
separated by several units of rapidity in $dAu$ collisions indicated
the possible onset of the "mono-jet" behavior expected in the quantum
CGC picture \cite{KLM1} (the azimuthal correlations in the classical
approach to the CGC were studied in \cite{KTaz}).

Nevertheless, the origin of the observed effects is certainly not
beyond a reasonable doubt at present; to clarify it, one needs to
perform dedicated and careful experimental and theoretical
studies. While the data is qualitatively consistent with the
predictions based on the CGC picture, a detailed comparison to the
data requires a quantitative analysis taking into account, for
example, the contributions of both valence quarks and gluons, and the
influence of realistic fragmentation functions. Such an analysis is
the goal of this note. Recently, related work in the more traditional
multiple scattering picture supplemented by shadowing has been done in
Refs. \cite{Accardi:2004ut,Barnafoldi:2004kh}, and in
Ref. \cite{Jalilian-Marian:2004xm} where the contribution of valence
quarks scattering off the CGC has been addressed.

In this work we use a simple model for the dipole--nucleus
forward scattering amplitude which describes the onset of the gluon
anomalous dimension in the Color Glass Condensate regime. 
 Since the inclusive gluon and quark production
cross sections in $p(d)A$ collisions can be expressed in terms of the
adjoint dipole--nucleus scattering amplitude, our model allows us to
describe inclusive hadron production in deuteron--gold collisions at
$\sqrt{s}=200$~GeV at RHIC. Our model is based on a detailed 
analytical analysis performed in our previous publication \cite{KKT}
stemming from the idea put forward in \cite{KLM}.

Inclusive cross section for production of a gluon in $dA$ collisions was
calculated in \cite{KM,KT,Braun} and is given by
\beq\label{paev}
\frac{d \sigma^{dA}_G}{d^2 k \ dy} \, = \, \frac{C_F \, S_A \, S_d}{\as \, 
\pi \, (2 \pi)^3} \, 
\frac{1}{{\un k}^2} \, \int d^2 z \, 
\nabla^2_z \, n_G ({\un z}, Y-y) 
 \, e^{- i {\un k}
\cdot {\un z}} \, \nabla^2_z \, N_G ({\un z}, y),
\eeq
where $S_A$ and $S_d$ are cross sectional areas of the gold and
deuteron nuclei correspondingly and $Y$ is the total rapidity
interval. We assume a simple form of the scattering amplitude of the
gluon dipole of transverse size $z_T = |{\un z}|$ on the deuteron
inspired by the two-gluon exchange
\cite{KKT}
\beq\label{deu}
n_G ( {\un z} , \ln 1/x_p) \, = \,(1-x_p)^4\, x_p^{-\lambda} \, \pi
\as^2 z_T^2 \ln (1/z_T \mu) \, \frac{1}{S_d}
\eeq
with $\lambda$ to be fixed later and $x_p$ the gluon's Bjorken $x$ in
the deuteron's (or proton's) wave function. Integrating over
directions of ${\un z}$ we rewrite \eq{paev} as
\beq\label{sigma}
\frac{d\sigma^{dA}_G}{d^2k\,dy}=\frac{\as C_F}{\pi^2}\frac{S_A}{k_T^2}
\, x_p^{-\lambda} \, \int_0^\infty dz_T \, J_0(k_T z_T) \, \ln\frac{1}{z_T\mu}
\, \partial_{z_T}[
z_T\,\partial_{z_T}N_G(z_T,y)],
\eeq
where $\mu$ is a scale associated with deuteron and is fixed at
$\mu=1$ $~GeV$ thereof. The gluon dipole scattering amplitude on a gold
nucleus $N_G(z_T,y)$ should be determined from the nonlinear evolution
equation \cite{BK}. Since an exact solution of the nonlinear evolution
equation \cite{BK} is a very difficult task we are going to construct
a model for $N_G(z_T,y)$ satisfying its asymptotic behavior: at
$z_T\ll 1/Q_s(y)$ one should have $N_G(z_T,y) \sim z_T^2$, while at
$z_T\gg 1/Q_s(y)$ we should get $N_G(z_T,y)\sim 1$
\cite{LT,IIM,BK}. ($Q_s(y)$ is the nuclear saturation scale at rapidity $y$.) 
This behavior can be modeled by a simple Glauber-like formula
\beq\label{glauber}
N_G(z_T, y)=1-\exp\left[-\frac{1}{4}(z_T^2 Q_s^2)^{\gamma(y,
z_T^2)}\right],
\eeq
where $\gamma(y, z_T^2)$ will be given by \eq{gamma}. Note, that when
$\gamma=1$ equations \eq{sigma} and \eq{glauber} reproduce the results
of McLerran--Venugopalan model \cite{MV,kjklw,KM} (for similar results
see \cite{AG}).

At forward rapidities, in the deuteron fragmentation region, the
Bjorken $x$ of the nucleus acquires its lowest possible value for a
given $\sqrt{s}$, while the Bjorken $x$ of the proton is close to
unity. In that region rescatterings of valence quarks of the proton in
a nucleus can give a substantial contribution to the hadron production
cross section. This problem was discussed in a series of papers listed
in \cite{qp,jd} leading to the following expression for inclusive valence
quark production cross section \cite{jd}
\beq\label{jamal}
\frac{d\sigma^{dA}_Q}{d^2k}=\frac{S_A}{2\pi}
\int_0^\infty dz_T \, z_T \, J_0(k_T z_T) \, [2 -  N_Q(z_T, y)],
\eeq
where $N_Q (z_T,y)$ is the quark dipole--nucleus forward scattering
amplitude. In the quasi-classical approximation ($\gamma$ =1) $N_Q (z_T,y)$ is
given by the same quasi-classical formula \eq{glauber} with $Q_s^2(y)$
replaced by $\frac{C_F}{N_c}Q_s^2(y)=\frac{4}{9}Q_s^2(y)$. Therefore,
by analogy with \eq{glauber}, we model the quark dipole scattering
amplitude $N_Q(z_T, y)$ as
\beq\label{glauber2}
N_Q(z_T, y) \, = \, 1- \exp\left[-\frac{1}{4} \left(z_T^2
\frac{C_F}{N_c} \, Q_s^2\right)^{\gamma(y,z_T^2)}\right].
\eeq

To model the anomalous dimension $\gamma(y, z_T^2)$ 
we use the following interpolating formula
\beq\label{gamma}
\gamma(y, z_T^2) \, = \, \frac{1}{2}\left(1+\frac{\xi 
(y, z_T^2)}{\xi (y, z_T^2) + \sqrt{2 \,\xi (y, z_T^2)}+  7
\zeta(3)\, c} \right),
\eeq
where 
\beq\label{xi}
\xi (y, z_T^2) \, = \, \frac{\ln\left[1/( z_T^2 \, Q_{s0}^2 ) 
\right]}{(\lambda/2)(y-y_0)}\,,
\eeq
and $c$ is a constant to be fitted. This form of the anomalous
dimension is inspired by the analytical solutions to the BFKL equation
\cite{BFKL}. Namely, in the limit $z_T \rightarrow 0$ with $y$ fixed we 
recover the anomalous dimension in the double logarithmic
approximation $\gamma \approx 1 - \sqrt{1/(2 \, \xi)}$. In another
limit of large $y$ with $z_T$ fixed, Eq. \eq{gamma} reduces to the
expression of the anomalous dimension near the saddle point in the
leading logarithmic approximation $\gamma \approx
\frac{1}{2} + \frac{\xi}{14 \, c \, \zeta (3)}$. Therefore Eq. \eq{gamma} 
mimics the onset of the geometric scaling region \cite{IIM,geom}. 
A characteristic value of $z_T$ is $z_T
\approx 1/(2 \, k_T)$, so we will put  $\gamma(y, z_T^2) \approx \gamma(y,
1/(4 \, k_T^2))$. 

The saturation scale $Q_s(y)$ that we use is the same as the one used in
\cite{GBW} to fit the low-$x$ DIS data and in \cite{KLN} to describe the 
hadron multiplicities at RHIC. It is given by
\beq\label{sat}
Q_s^2(y)= \Lambda^2\, A^{1/3}\, e^{\lambda y}= 0.13\, \mathrm{GeV}^2\,  
e^{\lambda\,y}\, N_\mathrm{coll}\,.
\eeq
Here $N_\mathrm{coll}$ is the number of binary collisions at a given
centrality in a $dAu$ collision. Parameters $\Lambda=0.6$~GeV and
$\lambda=0.3$ are fixed by DIS data \cite{GBW}. The initial saturation scale used
in \eq{xi} is defined by $Q_{s0}^2=Q_s^2(y_0)$ with $y_0$ the
lowest value of rapidity at which the low-$x$ quantum evolution
effects are essential.

The Cronin effect \cite{Cronin} is usually attributed to multiple
rescatterings of partons in the nucleus
\cite{AG,KNST,BKW,KKT}. However, it is also present in the low energy
data, i.e.\ at energies where saturation is unlikely to play a significant
role for the production of high $p_T$ particles. For example, at $\sqrt{s}=20$~GeV the nuclear enhancement
for $\pi^\pm$ produced in proton-nucleus collisions peaks at $k_T\simeq 4$~GeV \cite{Cronin}.  This implies
that the typical \emph{nonperturbative} scale $\kappa$
associated with such low energy hadronic rescatterings may be rather large. It
becomes much smaller than $Q_s(y)$ at high energies/rapidities as
one can see from \eq{sat}. However, at the central rapidity region at RHIC 
the influence of this non-perturbative scale cannot yet be neglected. 
 To take it into account in describing
the nuclear modification at RHIC we shift the saturation scale in
the Glauber exponents \eq{glauber} and \eq{glauber2} as follows
$Q_s^2\longrightarrow 
Q_s^2+\kappa^2\,A^{1/3}$.
This shift is also performed for $Q_{s0}^2$ in \eq{xi}. In our
numerical calculation we chose two values of $\kappa$: $\kappa=1$~GeV
takes into account additional momentum broadening due to a
nonperturbative effects and $\kappa=0$ neglects such effects.

The nuclear modification factor is usually defined as
\beq\label{rda}
R_{dAu}(k_T,y)=\frac{\frac{dN^{dAu}}{d^2k\,dy}}
{N_\mathrm{coll}\frac{dN^{pp}}{d^2k\,dy}},
\eeq
where $\frac{dN^{dAu}}{d^2k\,dy}$ and $\frac{dN^{pp}}{d^2k\,dy}$ are
multiplicities of hadrons per unit of phase space in dAu and pp
collisions.  Both expressions for gluon \eq{sigma} and quark
\eq{jamal} production contribute to the hadron production cross section 
in $dAu$ collisions. The cross section of hadron production is given
by
\begin{eqnarray}\label{hadr}
\frac{d\sigma^{dA}_h}{d^2k\,dy}& =&
\int \frac{dz}{z^2} \, \frac{d\sigma^{dA}_G}{d^2k\,dy}(k_T/z) \,
D_{frag}^G (z, k_T) \, F(k_T/z,y) \nonumber\\ &&+\int
\frac{dz}{z^2} \, \frac{d\sigma^{dA}_Q}{d^2k}(k_T/z) \,
xq_V(y,k_T/z)\, D_{frag}^Q (z, k_T) \, F(k_T/z,y). 
\end{eqnarray}
We use the LO fragmentation functions from Ref.~\cite{frag}. We choose
the renormalization scale of the fragmentation functions to be $k_T$.
Equation \eq{jamal} is derived for production of a valence quark in the 
deuteron fragmentation region. To generalize it to smaller values of 
Bjorken $x$ one has to convolute it with the deuteron's valence quark 
distribution, which is fixed by quark counting rules at high $x$ and by the 
leading Regge trajectory at low $x$
\beq\label{qv}
xq_V(x)=1.09\,(1-x_p)^3\,x_p^{0.5}\,,
\eeq
where $x_p=(k_T/\sqrt{s})\,e^\eta$.
(Eq.~\eq{qv} is normalized to give the distribution of a \emph{single} 
valence quark in the deuteron to keep normalization the same as in 
\eq{paev}.)  Valence quarks are increasingly
less important at low $x$ \cite{IKMT}, where the quark production is
dominated by gluons splitting in $q\bar q$ pairs. The factor of
$x_p^{0.5}$ insures that this is indeed the case here \cite{IKMT}.
Analogously, the high $x$ behavior of the \emph{nuclear} gluon
distribution is taken into account by introducing the function
$F(k_T,y)$
\beq\label{kinemat1}
F(k_T,y)=(1-x_A)^4\,
\left(\frac{\Lambda^2}{k_T^2+\Lambda^2}\right)^{1.3\,\as}\,.
\eeq
where the Bjorken $x$ of a gluon in the nuclear wave function is given
by $x_A=(k_T/\sqrt{s})\,e^{-\eta}$ and $\as =0.3$.  The last factor in
Eq.~(\ref{kinemat1}) arises when we impose momentum conservation
constraint on the anomalous dimension of the distribution
functions. Namely, we use the following phenomenological
parametrization of the anomalous dimension in the Mellin momentum variable
$\omega$ \cite{EKL}
\beq\label{mellin}
\gamma(\omega)=\as\, \left(\frac{1}{\omega}\,-\,1\right) \,.
\eeq
This parametrization takes into account high $x$ corrections to the QCD 
splitting functions.

The differential hadron multiplicity can be calculated by dividing
\eq{hadr} by the total inelastic cross section $\sigma_\mathrm{dAu}$
for a given centrality selection.  The baseline pp multiplicity is
calculated by expanding the Glauber exponent \eq{glauber} to the
leading term at $z_T \ll 1/Q_s$.  The free parameters of our model are
$y_0$ in \eq{xi}, which sets the initial value of $y$ at which the
quantum evolution sets in, $c$ in \eq{gamma}, which describes the
onset of quantum regime, the momentum scale $\kappa$, which specifies
the typical hadronic rescatterings momentum, and $\mu$ in \eq{deu},
which is the infrared cutoff. The value of $\mu=1$~GeV and the range
of values for $\kappa = 0 \div 1$~GeV are fixed by lower energy
data. Parameters $y_0$ and $c$ are fitted to RHIC $dAu$ data reported
by BRAHMS collaboration \cite{Debbe,BRAHMSdata}. The parameter
$\Lambda$ from \eq{sat} is fixed by the DIS data and is not a free
parameter in our model.

The data reported in Ref.~\cite{BRAHMSdata} is for charged particles
at pseudo-rapidities $\eta=0,1$ and for negative ones at
pseudo-rapidities $\eta=2.2,3.2$. At forward rapidities (in the
deuteron fragmentation region) the valence quarks begin to dominate
over gluons in the production of hadrons with high transverse
momenta. In particular, in $pp$ collisions this leads to an asymmetry
between positive and negative hadrons -- an effect which is
well--established (see \cite{iso} and references therein). Since the
nuclear modification factor $R_{dA}$ has been experimentally defined
as the ratio of $dAu$ and $pp$ cross sections, this factor is modified
by the isospin asymmetry effects. Unfortunately, it is difficult to
evaluate quantitatively the magnitude of these effects -- the isospin
dependence of fragmentation functions is poorly known, and the
relative importance of valence quarks and gluons in various
kinematical regions heavily depends on the choice of the structure
functions. Nevertheless, to account for the influence of this effect
we performed calculations for two limiting cases: (i) assuming no
isospin dependence for the valence quark fragmentation and (ii) in the
opposite limit of the constituent quark model, with $u$-quarks
fragmenting only into positive hadrons and $d$-quarks fragmenting only
into negative ones.

\begin{figure}
\begin{tabular}{cc}
\epsfig{file=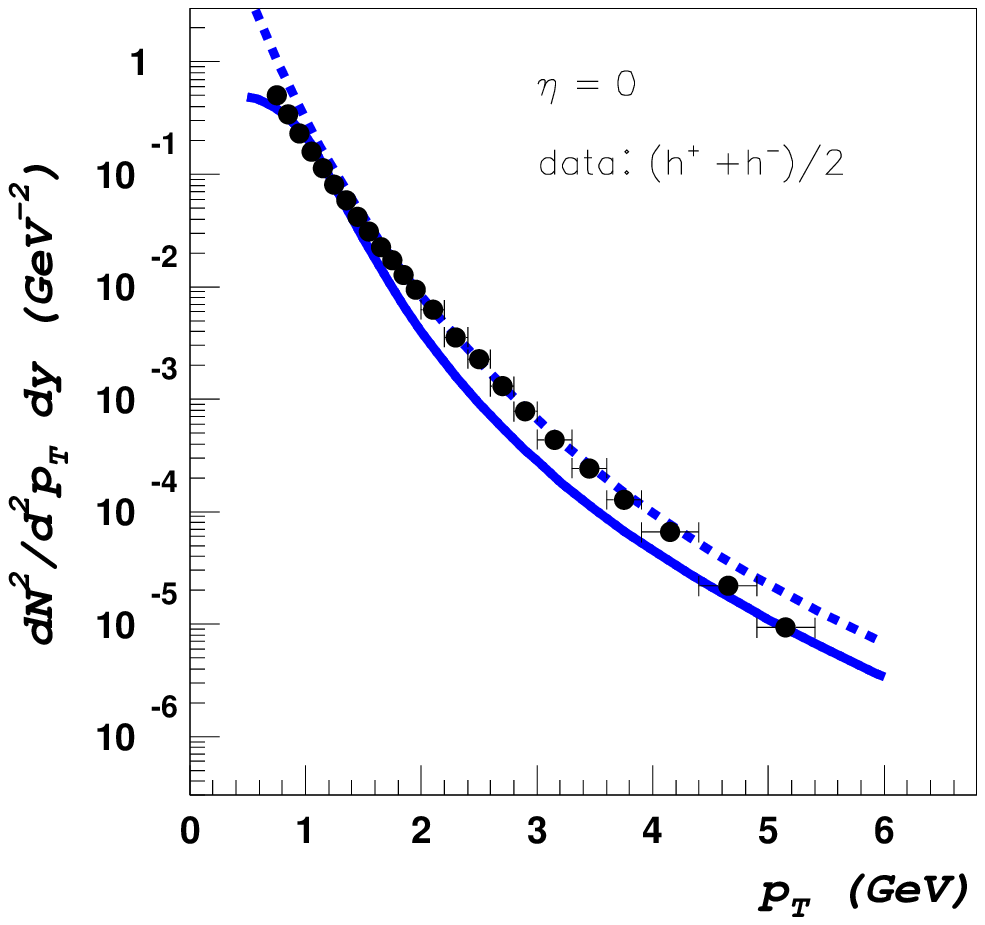,width=6.5cm}&
\epsfig{file=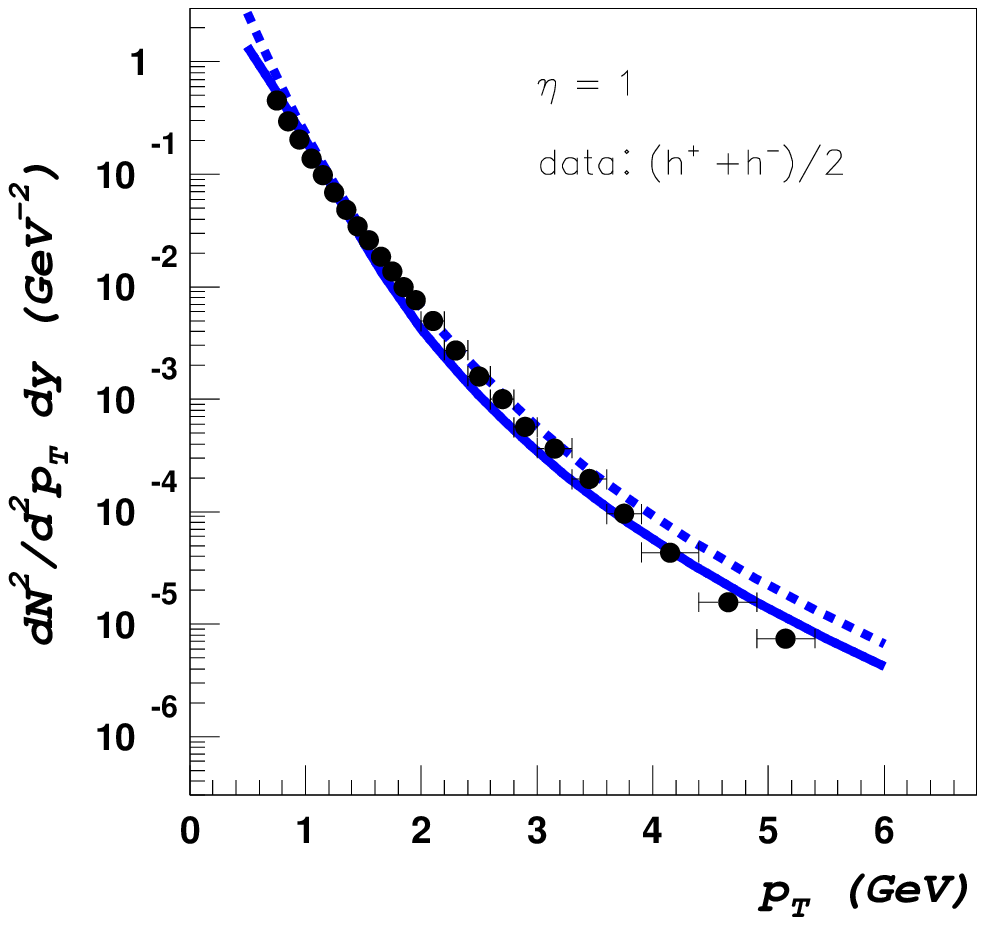,width=6.5cm}\\
\epsfig{file=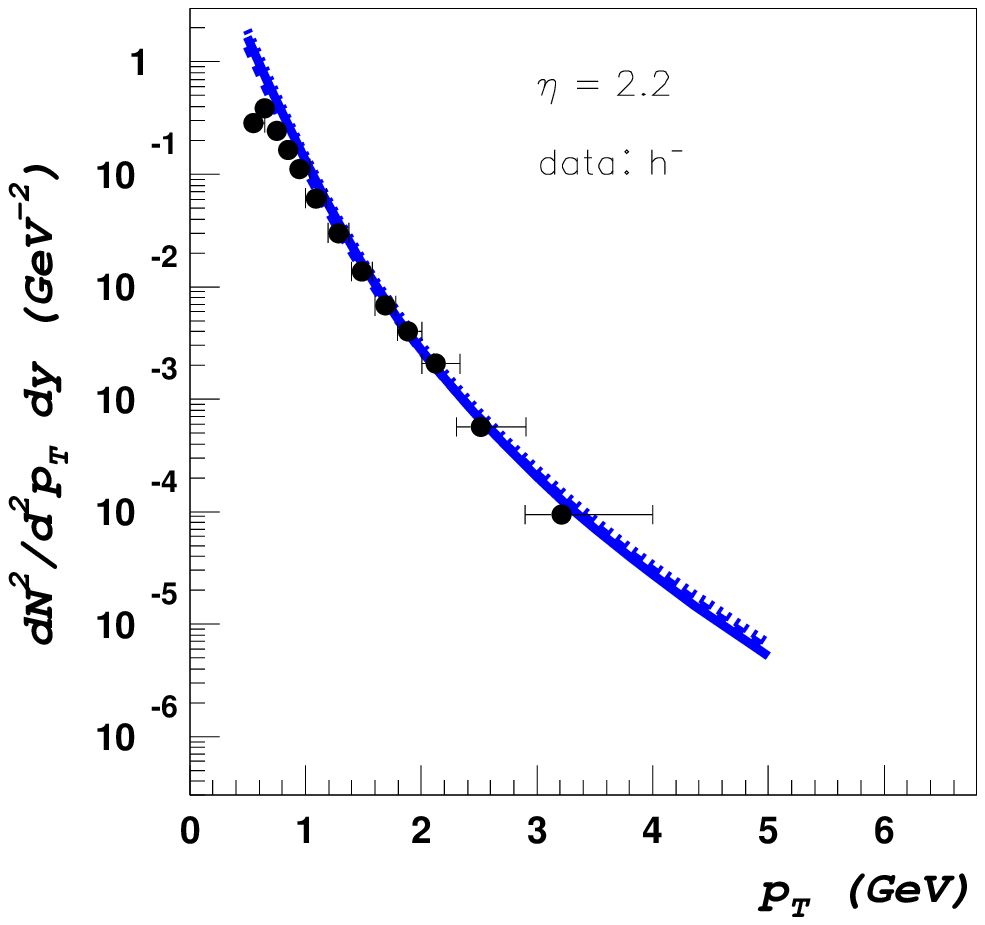,width=6.5cm}&
\epsfig{file=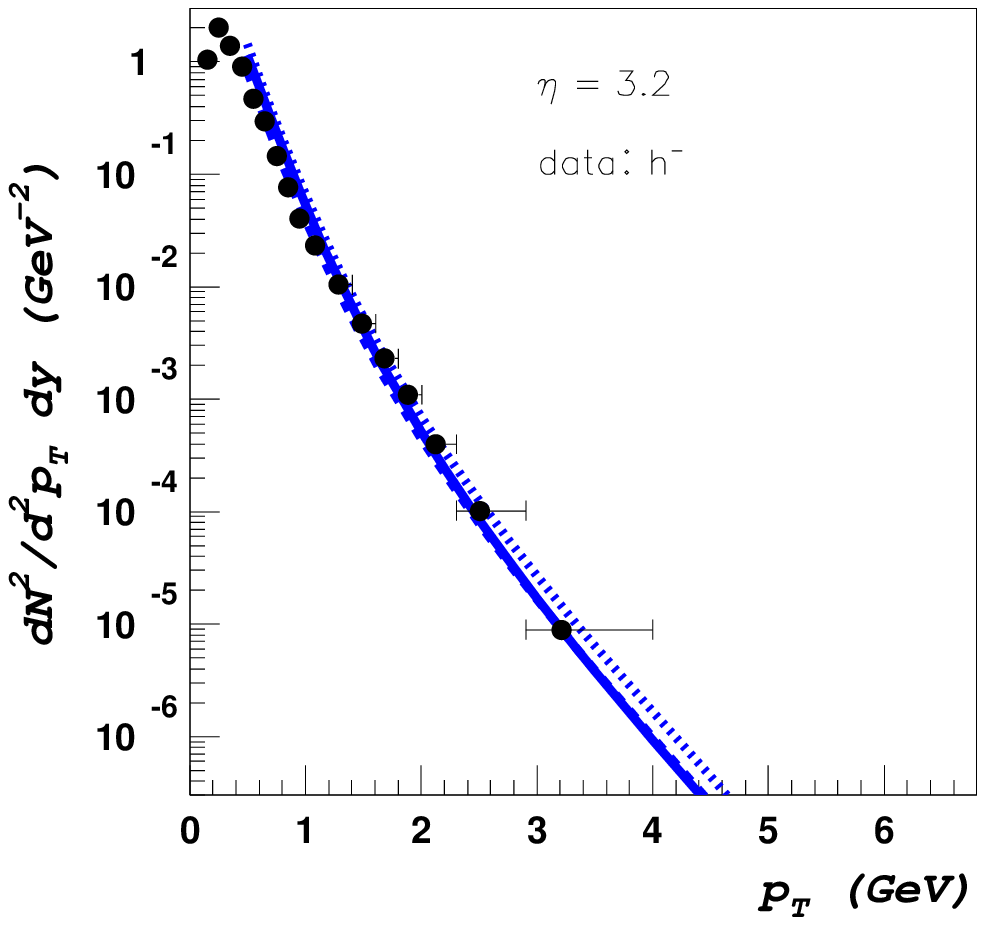,width=6.5cm}
\end{tabular}
\caption{Charged particle spectra in Deuteron-Gold 
collisions at $\sqrt{s}=200$~GeV at RHIC. For the plots with
$\eta=0,1$ the solid line corresponds to $(h^-+h^+)/2$ contribution
calculated in the isospin-independent approximation for the
fragmentation functions with $\kappa=0$, while the dashed line gives
the same $(h^-+h^+)/2$ contribution for $\kappa=1$~GeV. In the plots
for $\eta=2.2,3.2$ the solid line denotes the $h^-$ contribution
calculated in the constituent quark approximation with $\kappa=0$, the
dashed line gives the same $h^-$ contribution for $\kappa=1$~GeV,
while the dotted line at $\eta=2.2,3.2$ gives the $(h^++h^-)/2$
isospin-independent contribution calculated for $\kappa=0$. Data is
from \cite{BRAHMSdata}.}\label{fig:spec}
\end{figure}

\begin{figure}
\begin{tabular}{cc}
\epsfig{file=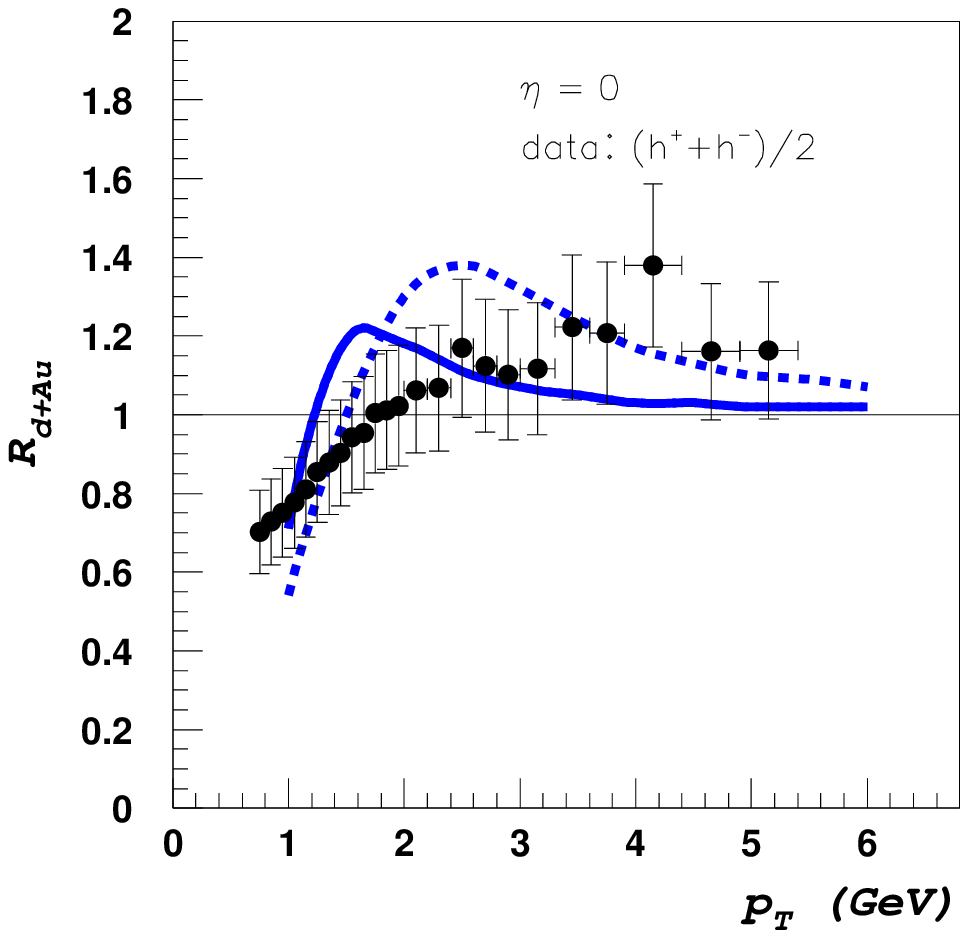,width=6.5cm}& 
\epsfig{file=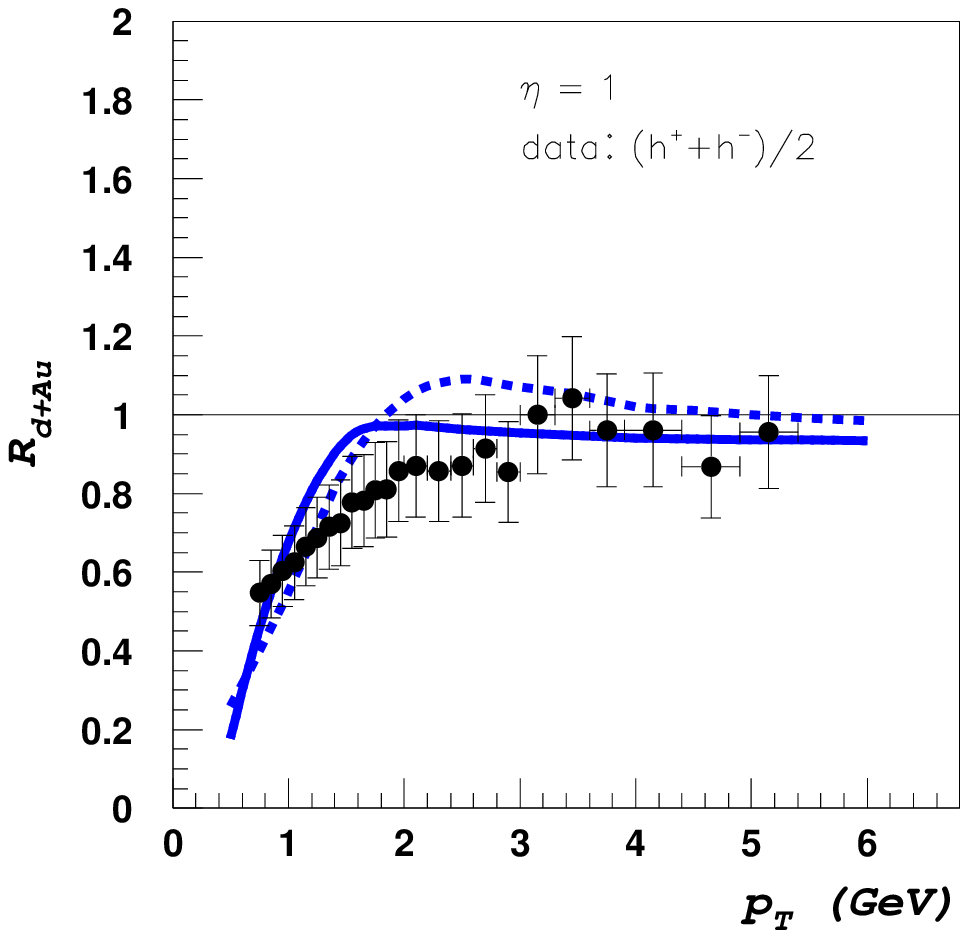,width=6.5cm}\\ 
\epsfig{file=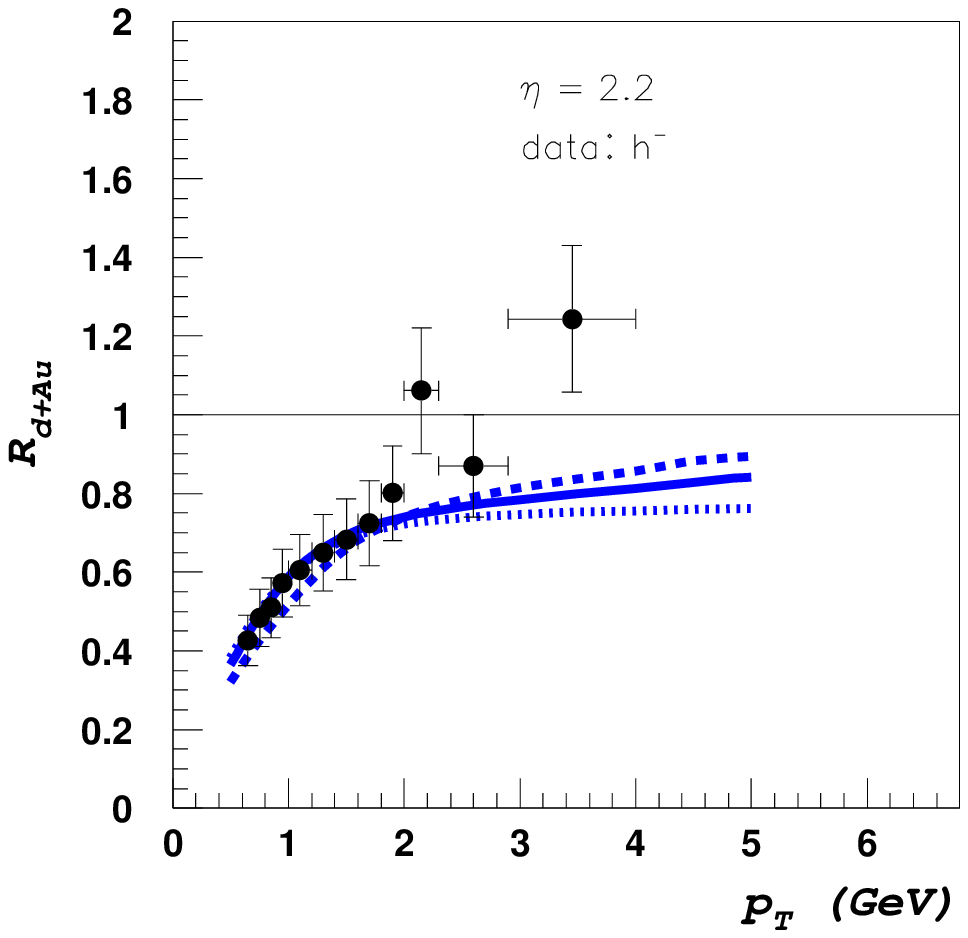,width=6.5cm}&  
\epsfig{file=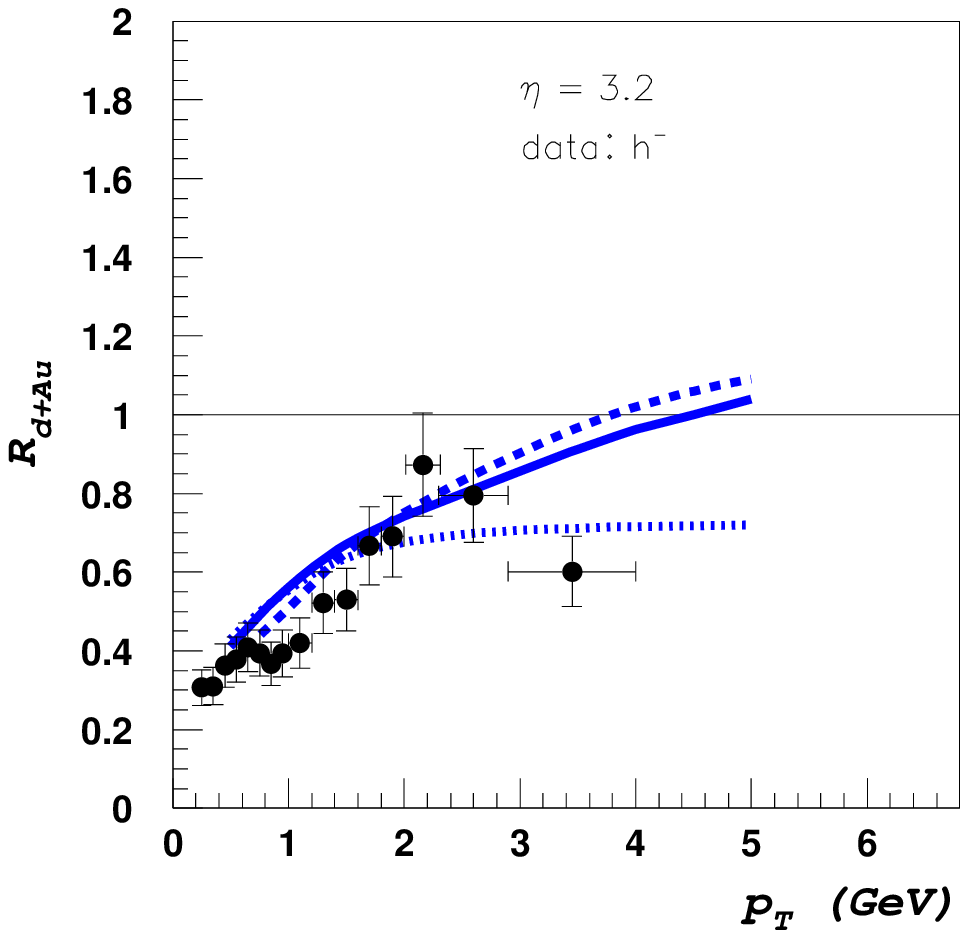,width=6.5cm}
\end{tabular}
\caption{Nuclear modification factor $R_{dAu}$ of charged particles 
for different rapidities. In the top two figures, corresponding to
$\eta=0,1$, the solid line corresponds to $(h^-+h^+)/2$ contribution
calculated with $\kappa=0$ in the isospin-independent approximation,
while the dashed line gives the same $(h^-+h^+)/2$ contribution but
with $\kappa=1$~GeV. In the lower two plots, corresponding to
$\eta=2.2,3.2$, the solid line gives the $h^-$ contribution calculated
in the constituent quark model with $\kappa=0$, the dashed line gives
the same $h^-$ contribution for $\kappa=1$~GeV, while the dotted line
at $\eta=2.2,3.2$ gives the $(h^++h^-)/2$ contribution with
$\kappa=0$.  Data is from
\cite{BRAHMSdata}.}\label{fig:rda}
\end{figure}

\begin{figure}
\begin{tabular}{cc}
\epsfig{file=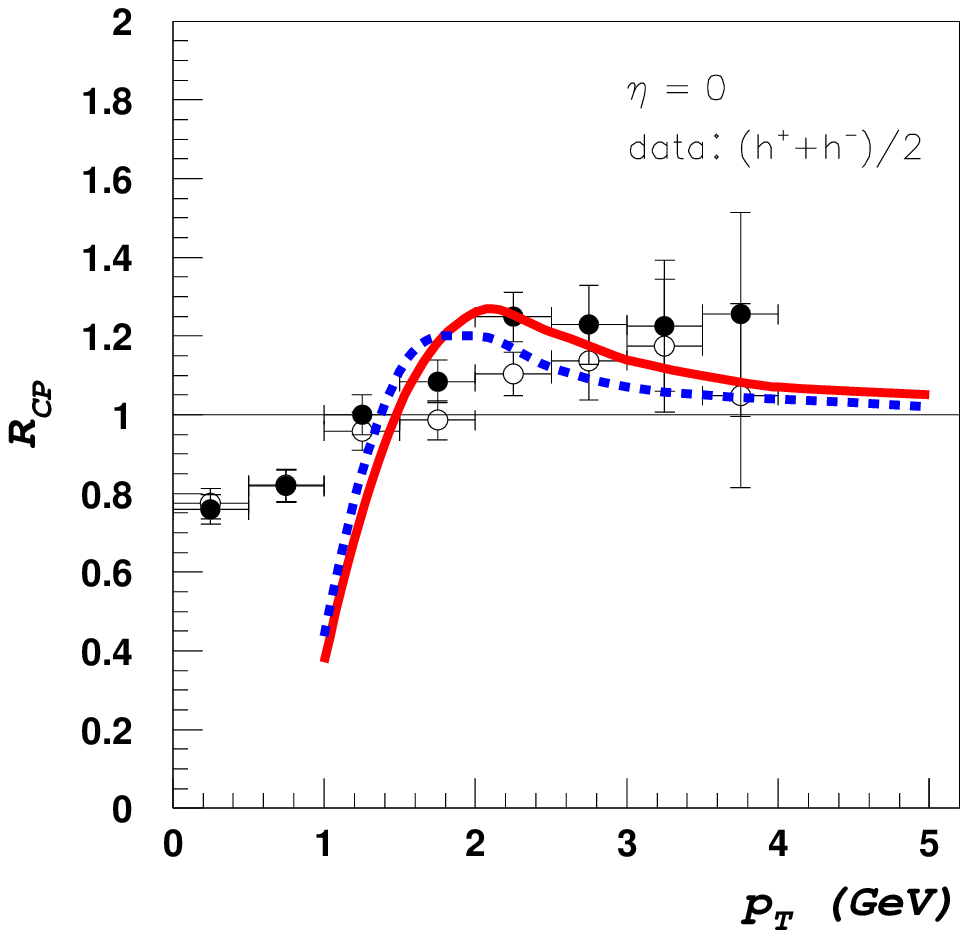,width=6.5cm}&  
\epsfig{file=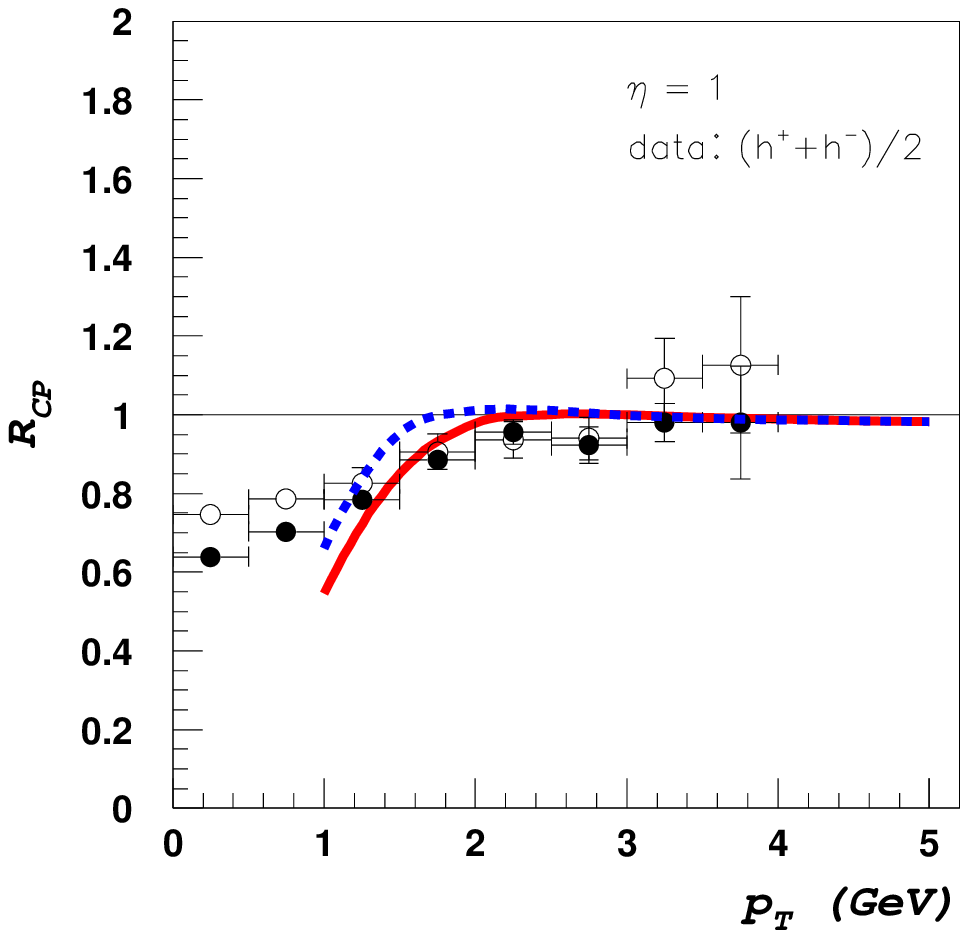,width=6.5cm}\\
\epsfig{file=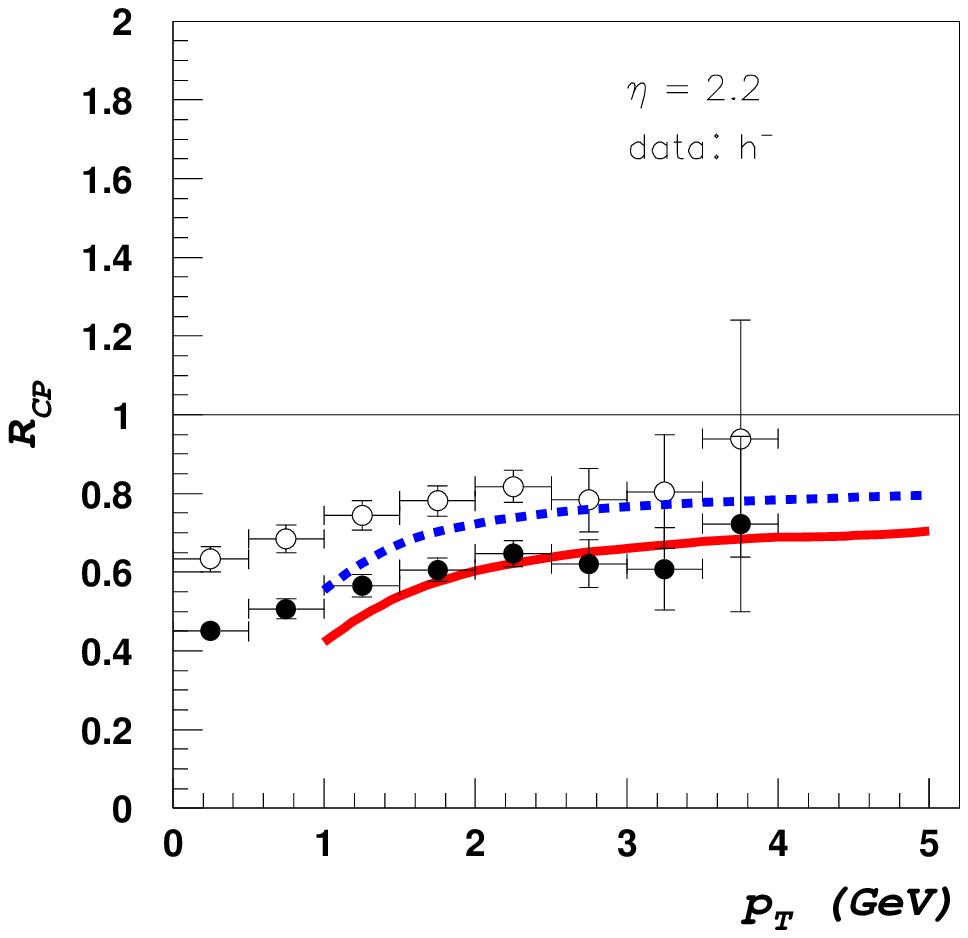,width=6.5cm}&
\epsfig{file=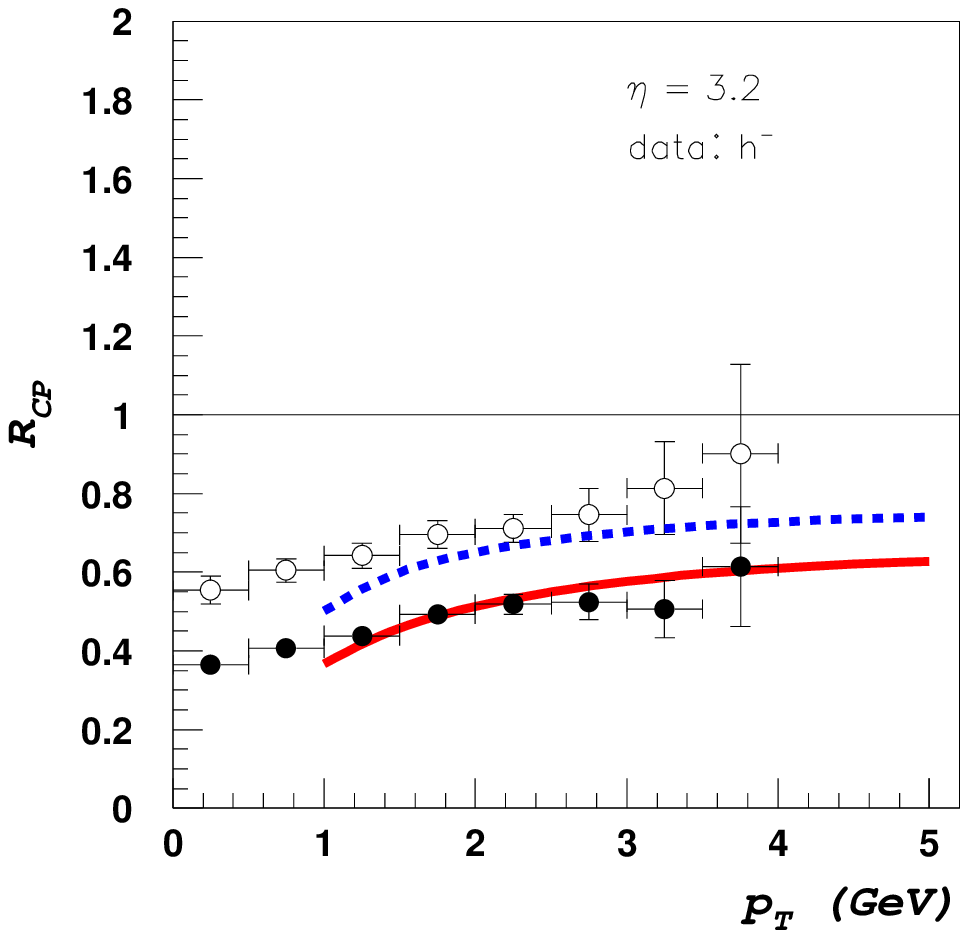,width=6.5cm}
\end{tabular}
\caption{Nuclear modification factor $R_{CP}$ of charged 
particles $(h^++h^-)/2$ calculated in the isospin-independent
approximation for rapidities $\eta=0,1$ and $R_{CP}$ of negatively
charged particles $h^-$ calculated in the constituent-quark model for
$\eta=2.2,3.2$ plotted for $\kappa=0$.  Data is from
\cite{BRAHMSdata}. Full and open dots, described by the solid and 
dashed lines correspondingly, give the ratio of particle
yields in 0-20$\%$ and 30-50$\%$ centrality events correspondingly
divided by the yields from 60-80$\%$ centrality events scaled by the
mean number of binary collisions \cite{BRAHMSdata}.}
\label{fig:rcp}
\end{figure}

The results of our calculations are presented in Figs. \ref{fig:spec},
\ref{fig:rda} and \ref{fig:rcp} along with the data collected by
BRAHMS collaboration \cite{Debbe,BRAHMSdata}. In these figures we use
$c=4$ with $y_0=0.6$ for both $\kappa=0$ and $\kappa=1$~GeV. We would
like to emphasize that the ratios $R_{dAu}$ and $R_{CP}$ are almost
insensitive to the values of $\kappa$ and $\mu$ at $\eta\ge 1$ and
$p_T\ge 2$~GeV. Their dependence on $y_0$ is also weak at forward
rapidities.

In \fig{fig:spec} we present our calculation of the charged particle
transverse momentum spectra in $dAu$ collisions at several different
rapidities compared to BRAHMS data \cite{BRAHMSdata}. We find a
reasonable agreement with experimental data \cite{BRAHMSdata}. To
evaluate the degree of agreement with the data one should also keep in
mind that BRAHMS data at $\eta = 0$ and $1$ are for the average of
positive and negative hadrons, and so contain the baryons which
production dynamics still remains puzzling at present (the shown data
at $\eta = 2.2$ and $3.2$ are for the negative hadrons only).  In
\fig{fig:rda} we show the nuclear modification factor $R^{dAu}$ as a
function of $p_T$ at different rapidities calculated in our model and
compared to the data from \cite{BRAHMSdata}. At rapidity $\eta < y_0$
we observe Cronin enhancement of the nuclear modification factor at
$p_T \sim 2 \div 3$~GeV due to the multiple rescatterings of the
deuteron in the gold nucleus \cite{KKT,AG,KNST,BKW}. At $\eta >y_0$
the low-$x$ quantum evolution effects modify the anomalous dimension
$\gamma$ leading to suppression in $R^{dAu}$ at forward rapidities and
disappearance of the Cronin maximum in accordance with our qualitative
predictions in Ref.~\cite{KKT} (see also \cite{Alb}).  \fig{fig:rcp}
demonstrates the centrality dependence of the hadronic spectra by
plotting the central-to-peripheral ratio $R^{CP}$ for the same
rapidities as in Figs. \ref{fig:spec} and \ref{fig:rda}. It is
important to emphasize that $R_{CP}$ is much less sensitive than
$R_{dAu}$ to isospin--dependent effects.  At mid-rapidity the Cronin
maximum increases and moves to the right as centrality
increases. Conversely, at forward rapidities, $\eta \gg y_0$, the
suppression gets stronger with centrality since, approximately,
$R^{dAu}\sim \sqrt{1/N_\mathrm{part}^{Au}}$
\cite{KLM,KKT}.  This behavior at forward rapidities is in agreement
with the BRAHMS data \cite{Debbe,BRAHMSdata}. Nuclear modification
factor obtained by numerical solution of the BK equation \cite{Alb} as
well as other approaches \cite{BKW,jd} qualitatively agree with our
conclusions. Further research in the area included an analysis of
running coupling corrections \cite{IIT} and a study of similar
suppression in di-lepton production \cite{BMS}.

\begin{figure}
\begin{center}
\epsfig{file=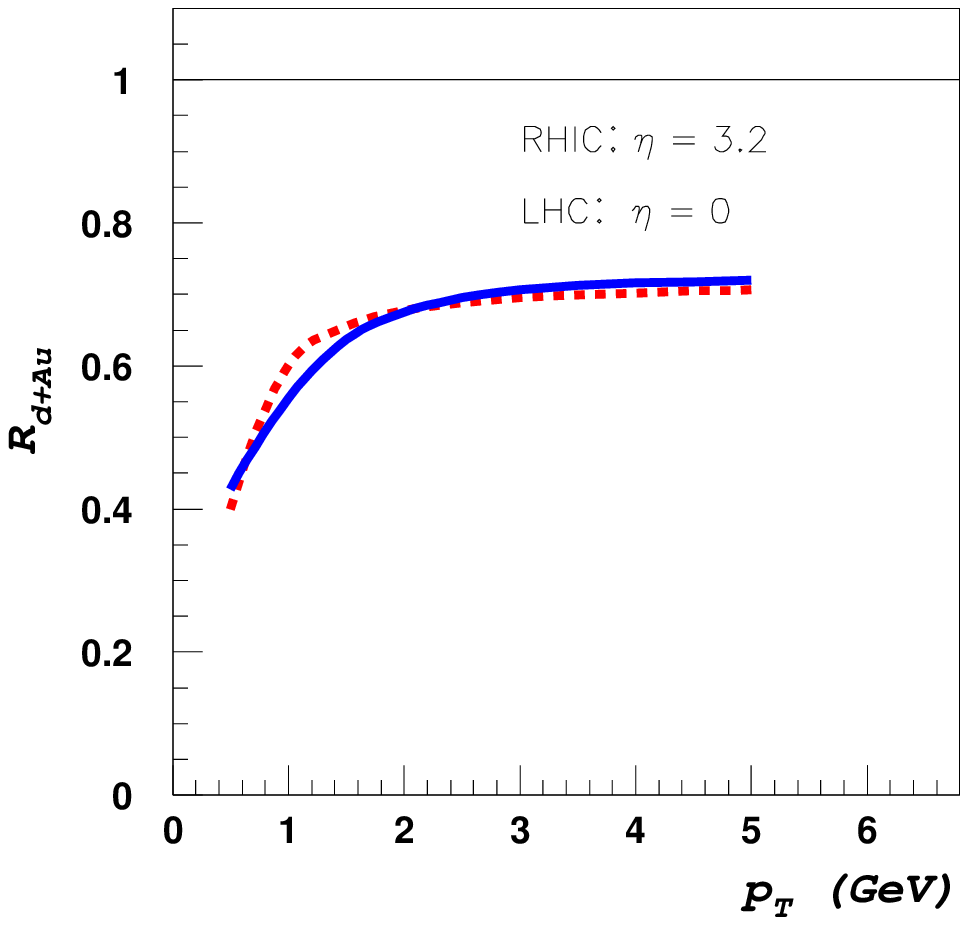,width=6.5cm}
\caption{Nuclear modification factor $R^{pA}$ of charged particles 
$(h^++h^-)/2$ at LHC energies $\sqrt{s}=5500$~GeV at mid-rapidity
$\eta=0$ (dashed line) versus $R^{dAu}$ of $(h^++h^-)/2$ for RHIC
energies $\sqrt{s}=200$~GeV at $\eta=3.2$ (solid line) plotted for
$\kappa=0$.}\label{lhc}
\end{center}
\end{figure}

It will be interesting to check the predictions of our approach at the
LHC energy of $\sqrt{s} = 5.5$ TeV. In \fig{lhc} we show our result
for the nuclear modification factor $R^{pA}$ for $pA$ collisions at
LHC at mid-rapidity, compared to $R^{dA}$ for RHIC $dAu$ collisions at
$\eta=3.2$. Our model predicts that the nuclear modification factor
will be quite similar for both cases. This conclusion seems natural to
us since the effective values of nuclear Bjorken $x$ for mid-rapidity
LHC collisions will be similar to the effective $x$ achieved in the
forward rapidity RHIC collisions. If observed, the mid-rapidity
suppression predicted in \fig{lhc} for $pA$ collisions at LHC would
indicate that an even stronger suppression due to the CGC initial
state dynamics should be present in $AA$ collisions at LHC. The
overall high-$p_T$ suppression in mid-rapidity $R^{AA}$ at the LHC
would then be due to both the initial state saturation/CGC dynamics
and jet quenching in quark-gluon plasma \cite{Bj,EL,BDMPS,EL2}.

In summary, we presented a simple but quantitative model which
incorporates the main features of the small-$x$ evolution in the Color
Glass Condensate for particle production in ultra relativistic proton
(deuteron)--heavy ion collisions. We find that at $\sqrt{s} = 200$ GeV
the evolution sets in at rapidity $y_0 \simeq 0.2$. As a result, at
central rapidities the nuclear modification factor $R^{dAu}$ exhibits
Cronin enhancement at $k_T\simeq 2 \div 3$~GeV, whereas at forward
rapidities it is strongly suppressed at all $p_T$.

\begin{acknowledgments}

We are grateful to Jamal Jalilian-Marian, Boris Kopeliovich, Eugene
Levin, Larry McLerran, and Werner Vogelsang for helpful discussions
and comments. The research of D.K. and K.T. was supported by the
U.S. Department of Energy under Contract No. DE-AC02-98CH10886.  The
work of Yu. K. was supported in part by the U.S. Department of Energy
under Grant No. DE-FG03-97ER41014.
\end{acknowledgments}


\end{document}